Title: Famed Bulgarian physicists. I. St. Petroff 's Göttingen research of the photostimulated
   interconversions of color centers in alkali halides: the discovery of the photostimulated
   aggregation
Author: Mladen Georgiev (Institute of Solid State Physics, Bulgarian Academy of Sciences,
   1784 Sofia, Bulgaria)
Comments: 8 pages with 3 figures and 1 picture, all pdf format
Subj-class: physics

This historical essay, intended for a wider audience, tells briefly of the life and work of one of the most successful scientists originating from a Balkan environment whose name and popularity have greatly exceeded its realm. The word is of a discovery during WWII of photo-stimulated aggregation of F centers (else alkali atoms) implanted from the vapor into an alkali halide crystal. Using optical absorption techniques while a grantee of Humboldt's Foundation in Göttingen, Germany between 1943-1944, he found new absorption bands pertaining to small-size F center aggregates and followed their interconversion. His observation had far-reaching consequences for two reasons: A primary photochemical solid state reaction was evidenced for the first time leading to products within the nanoscale range, now considered to be state of the art for modern science and technology.

1. Preface

Undoubtedly St. Petroff was one of the most renowned Bulgarian scientists. His popularity rising enormously after R.W. Pohl's publication of an account of Petroff's experimental work at Göttingen during WWII [1], he subsequently turned out to be an all time most cited Bulgarian scientists. Petroff's Göttingen work has laid down the foundations of a new branch of solid state physics, the one of *Photostimulated Aggregation of Color Centers*. The F centers distorting the lattice a bit more than the atomic-size, their lower aggregates certainly exceed the atomic framework to enter the nanoscale range, considered to constitute a state-of-the-art interest for modern science and technology.

2. Academic years

Stoyan Christov Petroff has been born on January 17, 1905 in the village of Polska Skakavitsa near Kyustendil in the southwestern part of the country, which had escaped luckily a quarter of a century before from the Ottoman Empire with the decisive effort of the Russian Imperial Army.

It is worth noting that much of the credit for justifying the liberation war is to be given to American diplomats and in particular to an American correspondent for the London Times, Januarius McGahan. In particular, McGahan could swing the public opinion against the Turkish atrocities following an unsuccessful uprising by the population in 1876. Emotionally albeit rightly, McGahan, who later cooperated with the Russian army, was posthumously called 'Liberator of Bulgaria.' But it has not been diplomats, missionaries, and journalists only, as knowledge-thirsty young Bulgarian population was educated by famed scholars at the

American College near Sofia (ancient Ulpia Serdica), the first American University to be founded outside of the United States, well over a decade before the official recognition of the newly formed Bulgarian State.

Petroff has graduated from a high school in Kyustendil in 1925 and during that same year he enrolled to Sofia University to study physics wherefrom he graduated in 1929. Upon graduation, he has been appointed immediately as an Assistant Professor in Physics. He has remained in that position until September 1932 when sacked along with many other colleagues because of budget deficiencies. From 1932 he has been a physics teacher in Plovdiv (ancient Philipopolis). During 1934 he has won, by competition, a teacher appointment for in the Third Model High School for Boys in Sofia where he remained until October 1945. We believe the above brief incursions into history, both ancient and more recent, may be helpful for the reader to orient himself.

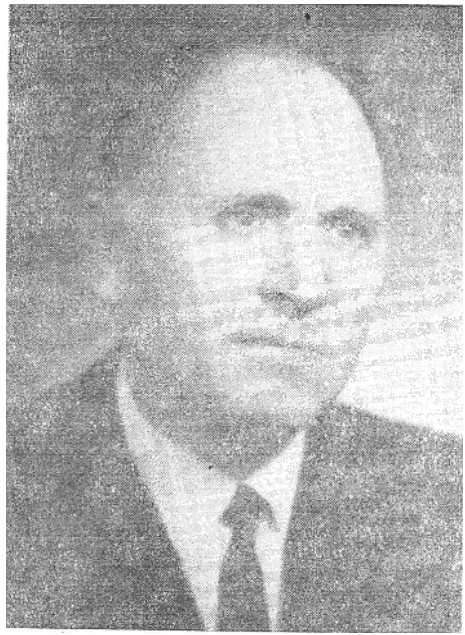

Picture 1. Petroff at the time of his seventieth anniversary in Rousse

3. Related physics at Göttingen

In 1942 Petroff has been granted a Humboldt Scholarship in Physics and has enrolled for research work at the First Physics Institute of the Göttingen University. For nearly two decades therein physicists under the supervision of the renowned scientist, Professor Robert Wichard Pohl, have been investigating the internal photoelectric effect and the optical properties of crystals [2]. The research activities at the Göttingen Scientific Center, now better known as Göttingen School, have founded wide areas of experimental research in solid state physics. They set forward many important branches, such as the photoelectric phenomena and the optical properties of solids. Besides their great fundamental significance, experimental results of the Göttingen School have an immediate impact on technology. In the latter respect we are tempting to mention the first in the historical retrospective development of the transistor based on colored alkali halides, one decade earlier than Shockley's semiconductor realization in the United States [3]. Perhaps the Göttingen workers have missed something the Americans possessed in abundance, the technical expertise based on a scientific vision. Nevertheless, the Göttingen School has given professional birth to a number of brilliant scientists, such as Pick, Stöckmann, and Stasiw (Germany), Gyulai (Hungary), Luty (USA), Savostianova (Russia) and so many others. Such was the environment Petroff enjoyed while in Göttingen. The scientific environment having left nothing more to desire, Petroff has rapidly found his place to start investigations of his own. During that time the state-of-the-art research problem for experimental physicists has been looking for analogies and closer relationship between the colored alkali halides and the photographic process in silver halides. In these investigations the alkali halides have been attached the role of a model medium.

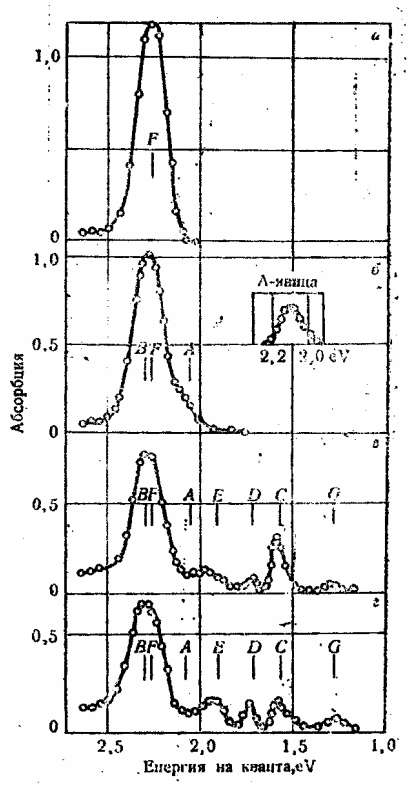

Figure 1. Petroff's sequence of bands: optical absorption vs. quantum energy (eV).

The problem lies in the following: the pure alkali halide crystal is colorless – it transmits non-selectively light from the whole visible part of the spectrum. But, under ionizing radiation, e.g. x-rays or γ-rays, the crystal is colored: sodium chloride colors yellow, potassium chloride turns violet, and so on. Virtually the same coloration can be achieved in the "additive way" by means of placing the crystal in the vapor of its metallic component at sufficiently high temperature – then the crystal dissolves a definite percentage of the vapor atoms. In all cases the color acquired by the crystal is due to the occurrence of an absorption band in the visible part of the spectrum, due to absorbing centers related to the dissolved metal atoms. The atomic structure of these centers is now firmly established: These are electrons substituting for the anions in the host lattice, or else, each center of that type is an electron trapped in the field of an empty anion vacancy. These vacancy centers have been named F centers by R.W. Pohl, and their related absorption band termed F band, from the German word *Farbzentrum*. The F center has a negligible mobility at room temperature. As a result, if a crystal is rapidly quenched from the additive coloration temperature down to 300 K it conserves the atomically dispersed state of its F centers formed at the coloration temperature. If now a freshly quenched crystal is illuminated with F band light at not too low a temperature, photo-aggregation processes run to form metallic colloid, a new phase of the metal dissolved. Here the photocoagulation process is formally akin to the photolysis in the silver halides, where prolonged illumination within the intrinsic absorption band leads to the formation of colloid silver particles within the crystal, the so-called print-out effect. In both cases the colloid formation leads to the occurrence of a new absorption band in the visible. Now, the question arises quite naturally just what the mechanisms and sequences are of the aggregation processes in both cases and whether there is any closer parallelism between them [4]. By virtue of Petroff's research conducted more than 60 years ago, many of these questions have been answered by himself, as well as by other authors, in relevance to the alkali halides, though not for the silver halides.

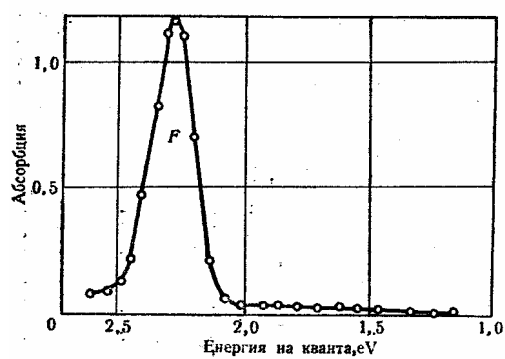

Figure 2. Optical absorption of the crystal with aggregate bands from Figure 1 after bleaching with wideband near-infrared light above 800 nm.

As it often happens, there have been predecessors to Petroff's photocoagulation work. Thus, Glaser and Lehfeldt [5] have reported that F centers formed by additive coloration may be made to coagulate when irradiated with white light at temperatures above the thermal stability range of F' centers. Apparently, Ottmer [6] has been the first to observe the aggregate M band, while Molnar[7] working independently from Petroff during WWII has undertaken a similar survey of the near-ifrared range where the F aggregate center bands occurred.

4. Photoaggregation

Examining cautiously the range between 600-1000 nm, Petroff observed the occurrence in a definite sequence of six new, well defined absorption bands which he designated by the letters A, B, C, D, E and G (Göttingen Nomenclature) in the order of their appearance. The occurrence of these bands at the consecutive stages of the illumination by F band light is shown in Figure 1. It was made clear later that the C, D, and E bands have been observed at nearly the same time by Molnar as well [7], while the G band has been rediscovered in 1949 by Burstein and Oberly [8]. Petroff's C band is Molnar's M band, first observed by Ottmer in 1928. In the American Nomenclature (nowadays accepted commonly), the long wavelength bands (except for the B band) are designated by the letters $F_{A1}$, M, $R_2$, $R_1$, and N. Seitz [9] proposed that the M band be designated in recognition of Molnar's discovery. During that WWII time Americans are aware of neither Petroff's discoveries nor of Ottmer's thesis.

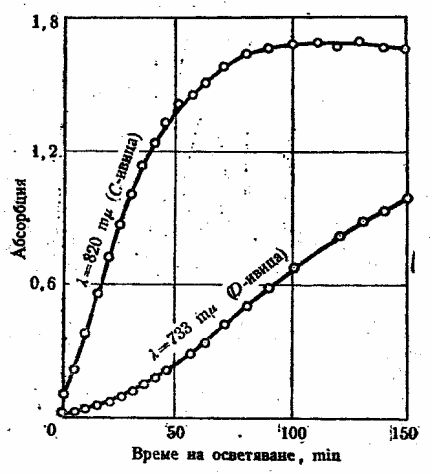

Figure 3. Sequential development of the C and D absorption bands upon illumination in the F band. The prior formation of the C band is necessary for the D band to start growing.

Contrary to his predecessors, Petroff not only evidenced the appearance of the absorption bands but he also established their genetic relationships, between themselves and with the F band. He showed that all the new bands occurred at the expense of the F band, since the area under the bands remained constant during the illumination. Every two bands of the F, A, B, C, D. E, G sequence have been in a paternal relationship between themselves in that each one of them grows at the expense of the preceding one. This can be seen convincingly in Figure 2, showing the growth curves of the C and D bands. It is clear that the D band starts growing at an appreciable rate only after the C band has reached a more considerable height. In other words, the availability of absorbing centers responsible for the C band is necessary for the D centers to start growing.

Another important discovery made by Petroff is the optical destruction (bleaching) of the long wavelength bands. Figure 3 displays a particular bleaching of the spectrum in Figure 1 by illumination with infrared light over 800 nm at 160°C. Of particular interest is the selective bleaching with light in the C band running easily even at room temperature.

Next in line of Petroff's discoveries has important implications for the general interpretation of the results obtained. He found that both the initial growth rate and the saturated magnitude of the long wavelength band C (cf. Figure 1) depended on the excitation light intensity. This indicates that a finite time elapses between the acts of light absorption by the F center and of the formation of the absorbing center responsible for the C band. Petroff called that time "Umbauezeit" (rebuilding time) and suggested a method for its experimental measurement. The method is in illuminating the crystal with chopped F light and investigating the dependence of the initial growth rate of the C band on the chopping rate. The rebuilding time determined in this way by Petroff has proved to be of the order of a second at room temperature. Obviously this new important point of Petroff's investigation indicates clearly that relatively slow ionic processes take part in the F→C interconversion.

The model interpretation is the generalizing step in Petroff's Göttingen work. The results obtained show clearly that the occurrence of the A, B, C, D, E, and G sequence of absorption bands upon illumination in the F band results from the running of definite photochemical reactions in the crystal. In Petroff's view the observed photochemical conversions have a clear meaning of photocoagulation which subsequently produces aggregates of two, three, and so on F centers upon illumination in the F band. The specific aggregation mechanism suggested by Petroff is, therefore, of the form:

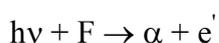

$h\nu + F \rightarrow \alpha + e'$

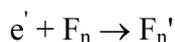

$e' + F_n \rightarrow F_n'$

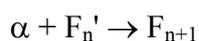

$\alpha + F_n' \rightarrow F_{n+1}$

where $F_n$ is an aggregate of n F centers, $F_n'$ is a charged intermediate product, formed as an $e'$ electron is trapped at $F_n$, and $\alpha$ is an anion vacancy. It is to be noted that the validity of the specific model proposed does not at all influence the heuristic value of the above-made general interpretation of the observed inter-conversions.

Perhaps it is worth noting that Petroff's mechanism of sequential photochemical aggregation of F centers is similar in principle to the respective mechanism proposed by Gurney and Mott to explain the formation of photolytic silver in the photographic process [9]. In their theory, a photolytic silver particle grows by way of subsequent steps of photoelectron trapping and neutralization of the trapped electron charge by adsorption of a mobile positive silver ion $Ag_i^+$ at the silver particle:

$$e' + Ag_n \rightarrow Ag_n'$$

$$Ag_i^+ + Ag_n' \rightarrow Ag_{n+1}$$

We fancy that Petroff in 1944 could not have read Gurney & Mott's paper published in England at the outbreak of hostilities leading to WWII. Nowadays Gurney & Mott's model reactions have the status of a solid state principle for photochemical decomposition of ionic salts by way of consecutive electronic and ionic multiple steps at definite sites in the crystal. Undoubtedly, a similar general validity can be attached to Petroff's model reactions as regards the photo-induced coagulation of F centers in that this coagulation proceeds by way of sequential running of electronic and ionic processes in the crystal.

## 5. Post factum

In 1946 the famed American scientist Frederick Seitz published a survey paper in Reviews of Modern Physics on the available theoretical and experimental investigations of alkali halides [10]. He devotes considerable space on the photochemical conversions in colored salts, though only on Molnar's studies. Apparently, Molnar has been engaged in wartime research of color center based screens for radar applications so that his work had little publicity before 1946. Seitz leaves no space to Petroff's investigations of which he has no prior knowledge. However, he qualifies Molnar's observations as photoaggregation, as Petroff has done. Seitz proposed models for the photochemical products which have much in common with Petroff's, though not all. In 1954 Seitz published a second review on the subject: now the attitude to Petroff's research is completely revised and it has been devoted a considerable space [11]. However, for his new outlook Seitz has received information from a source other than Bulgarian.

In 1950 a detailed account of Petroff's Göttingen work has been published on his behalf by Heinz Pick and Robert W. Pohl in the leading German journal Zeitschrift für Physik [1]. In a brief introduction they considered the background and pointed out the exact period of the investigation: 1943-1944. This paper has immediately attracted considerable interest among scientists the world over, from Europe to Japan and the United States, it has been cited in monographs and shorter articles on the colored crystals. The peculiarities of its publication have incited the admiration for the honesty of its publisher, the great German scientist R.W. Pohl. All in that same spirit, R.W. Pohl presents a paper at a symposium in England on the theory of the photographic process entitled "Report on Petroff's works on the photochemistry of crystals of potassium chloride." [12] The paper has been translated into Russian as well and published in a special volume, despite of their original skepticism. Petroff could not have dreamed to attend the symposium organized in his honor. At that time he was an Associate Professor in Physics at Varna University [13], though he had a "tail" as being pro-Western minded.

6. Epilogue

In 1954 The Varna University has been disbanded and Petroff was sent to Rousse on the Danube to head, as a regular professor, the Physics Department at the newly opened VIMMESS Agriculture Mechanization Institute. Therein he was engaged in research and teaching until his demise on July 12 1991. Before that he has left much of his Göttingen heritage to his younger colleagues and followers.

A small country has been blessed to have a scientist of Petroff's rank but could not stomach it. On his way back to the home country he has met skepticism: "Who do you think you are to claim that you have made a discovery of that significance?", his boss reacted emphatically. The pro-Soviet rulers were even more practical. A Western minded democrat would have been too dangerous to keep in teaching position at the capital. Accordingly, they sent him to the country, in a remote town far from libraries and scientific communities, a real dead-end street. The best way to kill a mind is to deprive him of his natural nourishing environment. And who do you think has been the loser? The country's physics and scientific self-confidence. Both were sacrificed in the name of Big Brother's global ambitions.